\documentstyle[aps,prl,floats,epsfig,preprint]{revtex}
\begin{document}
\draft

\title{Peak-effect and surface crystal-glass transition for surface-pinned
vortex array}

\author{
B. Pla\c cais, N. L\"utke-Entrup, J. Bellessa, P. Mathieu, Y.
Simon, and E.B. Sonin$^1$}
\address{Laboratoire de Physique de la Mati\`ere Condens\'ee de l'Ecole
Normale Sup\'erieure, \\
  CNRS-UMR8551, 24 rue Lhomond, 75231 Paris,
France
\\ $^1$Racah Institute of
Physics,  Hebrew University of Jerusalem, Jerusalem 91904, Israel}

\date{\today} \maketitle

\begin{abstract}
The peak effect has been investigated in clean Nb crystals with
artificially corrugated surfaces by measuring the linear surface
impedance in the 1kHz-1MHz frequency range.  From a two-mode
analysis of the complex spectra, we establish that vortex dynamics
is governed by surface pinning and deduce the associated vortex
slippage length. We demonstrate experimentally and theoretically
that the peak effect is related to a transition from collective to
individual surface pinning. A proper account of the peak-effect
anomalies implies softening of the shear rigidity by
disorder-induced lattice deformations. This leads to a vortex
crystal-glass transition induced by surface defects.

\end{abstract}

The peak effect (PE) is a well known anomaly of the mixed state
due to vortex lattice (VL) softening in the presence of random
defects \cite{Pippard69,Larkin}. It is observed at the upper
critical field $B_{c2}$ of clean superconductors
\cite{Autler62,Kes81},  well below $B_{c2}$ in disordered oxide
superconductors (second peak), or at the vortex-melting line in
very clean YBa$_2$Cu$_3$O$_7$ crystals \cite{Shi99}. The physics
of PE is the {\em order-disorder transition}, which is of interest
for a broad range of elastic media. Recent developments concern
vortices in conventional materials (Nb, NbSe$_2$), where PE is
investigated in great detail \cite{Paltiel00,Gammel98,Ling01}.
Originally believed to be a crossover from collective to
individual vortex pinning according to the Larkin-Ovchinnikov (LO)
scenario \cite{Larkin}, PE is now considered as a genuine
transition \cite{Paltiel00,Ling01}, with a quasi-long-range order
(Bragg-Glass) in the low-field phase \cite{Giamarchi95,Klein01}.
However, strong limitations for a quantitative analysis of the
phenomenon arise from the poor knowledge about the random defects
responsible for PE.

In this work we consider a new situation where this difficulty can
be overcome : the peak effect due to the {\em surface pinning of
vortices}. The pinning landscape is provided by the surface
corrugation $\zeta(\bf{r})$ of a slab in perpendicular or tilted
magnetic field; a controlled roughness $\zeta^*\!\sim1$-$10$nm is
produced by ion beam etching (IBE) and measured by atomic-force
microscopy (AFM). Far from the upper critical field $B_{c2}$ the
intervortex interaction is strong and suppresses the surface
pinning. This means that the surface pinning is {\em collective}
similar to the collective pinning in the bulk \cite{Larkin}.  But
close to $B_{c2}$ the VL distortion by surface pinning become
essential, and near the surface the ordered VL transforms to the
disordered vortex array, which we call {\em surface vortex glass}.
Formation of the surface vortex glass is accompanied by the
crossover from collective to individual pinning. This results in
the growth of the critical current, i.e. in PE.

Experiments are performed on a series of millimeter-thick slabs of
very pure Nb. Starting from electro-chemically polished samples, a
controlled corrugation is introduced by ion-beam etching (IBE) the
surfaces with a 1.5mA/cm$^2$ flux of 500eV-Ar$^+$ ions for
$10$--$100$ min. Given the sputter yield, $0.6$ Nb atom per
Ar$^+$, and the atomic spacing $c=0.26$ nm, the average etching
rate is of the order of 100 nm/min. From this stochastic process,
one expects a roughness amplitude $\zeta^*\simeq(c\overline{\Delta
z})^{1/2}$ and a white spectrum $S_{\zeta}({\mathbf k})=\int
{\mathbf dr}\,e^{-i\mathbf k \mathbf r}\langle \zeta({\mathbf
r}+{\mathbf R})\zeta({\mathbf R})\rangle_{\mathbf R}
=c^3\overline{\Delta z}/\pi$ in the range $0-c^{-1}$ of wave
number $k$. In practice, the wave number cut-off  $k_c <1/c$ due
crystal rearrangement associated with local heating; whence a
reduction of $\zeta^*$ by a factor $k_c c\!\sim\!10^{-1}$. This
picture is confirmed by AFM-inspection, which shows a flat
spectrum below $1\mu$m$^{-1}$ and a power law dependence
$S_{\zeta}(k)\!\sim\!1/k$--$1/k^2$ above. The etched surfaces have
a  broad-band corrugation, in the range $k=1$--$100 \mu$m$^{-1}$
covering the vortex reciprocal-lattice unit
$Q\!=\!a_0^{-1}\!\simeq\! (50$nm)$^{-1}$, which is very effective
for VL-pinning. Here $a_0=\sqrt{\varphi_0/\pi B}$ is the radius of
the vortex Wigner-Seitz cell (on the order of the intervortex
spacing) and $\varphi_0=h/2e$ is the flux quantum.

The pinning strength is deduced form the AC response. The AC
surface impedance $Z(\omega)=-i\mu_0\omega \lambda_{AC}$ is
determined by the effective penetration depth   given by
\begin{equation}
\frac{1}{\lambda_{AC}}=\frac{1}{L_S}+\left(\frac{1}{\lambda_{C}^2}+i\omega
\mu_0\sigma_f\right)^{1/2};\quad
L_S=\frac{l_SB}{\mu_0\varepsilon}\quad .
\label{two_mode}\end{equation} Here $\sigma_f$ is the flux-flow
resistivity, $\varepsilon\varphi_0$ is the vortex-line tension
(energy per unit length), and $\lambda_C$ the Campbell penetration
depth for bulk pinning. The surface pinning is taken into account
by the length $L_S\!\sim\!0.1$--$100 \mu$m, which can simulate the
effect of bulk pinning characterized by the Campbell depth
$\lambda_C$. This expression was derived within the frame of the
{\em two-mode electrodynamics} \cite{Sonin92,Entrup}, which
incorporates the surface pinning by introducing a phenomenological
boundary condition,
\begin{equation}
\varepsilon\varphi_0\left({{\mathbf u}(0) \over l_S}+  {\partial
{\mathbf u}(0) \over \partial z}\right)=0~,
     \label{BC}\end{equation}
imposed on the average VL displacement ${\mathbf
u}(z)=\overline{{\mathbf u}({\mathbf r},z)}$  at the surface of
the sample, which occupies the semispace $z<0$. Here $l_S$ is a
{\em slippage length} and the displacement ${\mathbf u}$ is
averaged over the position vectors ${\mathbf r}$ in the $xy$
plane. Physically this boundary condition presents the balance
between the pinning force $-\varepsilon\varphi_0 {\mathbf u}(0) /
l_S$ and the line tension force $\varepsilon\varphi_0 \partial
{\mathbf u}(0) /\partial z$ due to the vortex bending.

\begin{figure}[!!!t]
\centerline{\epsfig{file=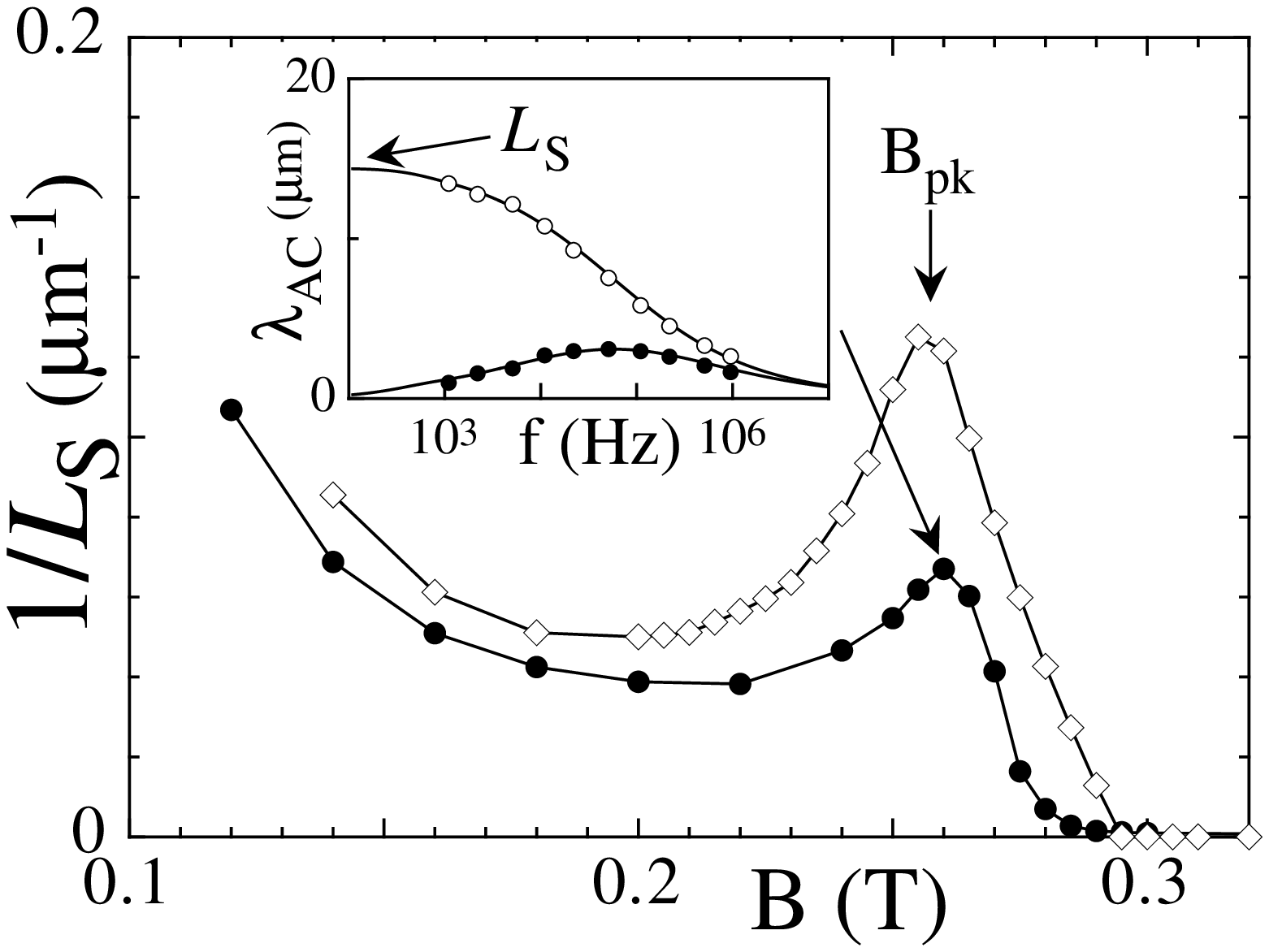, scale=0.4}}
\caption{Peak-effect in the elastic response $L_S^{-1}(B)$ of a
surface-pinned vortex array. Diamonds and full circles correspond
to oblique and perpendicular field orientations in sample $\#$S2
at T=4.2K. Inset:  the frequency dependence of the real and
imaginary parts (open and full circles respectively) of the
penetration depth $\lambda_{AC}(B,f)$ from which $L_S$ is deduced.
Solid lines are theoretical fit to Eq.(\ref{two_mode}) with
$\lambda_C=\infty$, $\sigma_f^{-1}=10$nOhm.cm and $L_S=14.9\mu$m.}
\label{pic_effect}
\end{figure}

Figure \ref{pic_effect} shows the PE due to surface pinning. Data
relate to sample $\#$S2 with dimensions
$25\!\times\!10.1\!\times\!0.87$ mm$^3$. The sample was annealed
in ultra-high vacuum  which gives a low residual resistivity
$\rho_{\rm n}\!=\!11$ n$\Omega$cm and $B_{\rm c2}\!=\!0.29$ T at
$4.2 K$. This value is consistent with earlier measurements within
the anisotropy  of Nb crystals ($B_{c2}(\theta)\!=\!0.280$-$0.295$
at 4.2K)\cite{Williamson70}. From the 90 min IBE exposure we
estimate $\zeta^*\sim 5$ nm, in qualitative agreement with AFM
measurement. Data points are obtained by fitting the
penetration-depth spectra as shown in the inset of the figure.
Metastability in the vortex number and/or arrangement is removed
by feeding a large transient current ($I\!=\!40$A) in the sample
prior to measurement. The abrupt onset of the AC-flux penetration
along the samples edges which are parallel to the field precludes
quantitative analysis for $B\gtrsim 0.95 B_{c2}$; this  difficulty
can be overcome by working in oblique field (figure
\ref{pic_effect}). More experimental details including the
calibrating procedure can be found in Refs.\cite{Entrup}. Already
present in pristine samples, the PE is strongly enhanced by IBE up
to a factor of 3 at the maximum dose.  This contrasts  with
conventional mechanical or chemical treatments which show
ineffective in increasing the PE in our Nb samples. This
distintive property of IBE emphasizes the importance of
small-scale atomic corrugation for the surface pinning of vortex
lines.

We separate quantitatively the bulk and surface contributions by
relying on the full 1kHz--1MHz spectrum $\lambda_{AC}(f)$
\cite{Entrup}.  From  fits of $\lambda_{AC}(f)$ with
Eq.(\ref{two_mode}) we deduce $L_S$, $\lambda_C$ and $\sigma_f$.
Remarkably enough, we always find $\lambda_C\!\gtrsim\!10$m,
meaning that the bulk pinning strength is negligible.  In fact
there is no difference in the spectra taken on both sides of the
peak which are equally well described by Eq.(\ref{two_mode}). This
confirms that surface pinning is most relevant for the vortex
dynamics in our experiment. The oblique-field data are larger by a
factor $\sim\!2$; this general trend is due to
surface-reinforcement of superconductivity in tilted fields.
Similar data are observed at lower temperatures, with, however,
larger $B_{pk}\!=\!0.95B_{c2}$ (1.8K) and a less pronounced PE.

\begin{figure}[!!!t]
\centerline{\epsfig{file=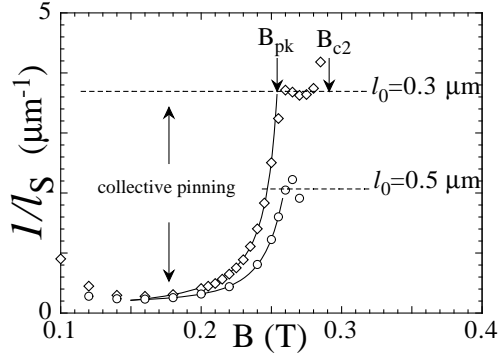, scale=0.4}}
\caption{The inverse of phenomenological slippage length $l_S(B)$
for a vortex array at a rough surface. It is deduced from the
$L_S$ data in Fig.\ref{pic_effect} according to the definition
$l_S=L_S\mu_0\varepsilon/B$ in Eq.(\ref{two_mode}). The effect of
vortex interactions in the collective-pinning regime below
$B_{pk}$ is visible as a strong reduction in $1/l_S(B)$. Solid
lines are power-law fits to the data at the transition.}
\label{slippage}
\end{figure}

Using the Abrikosov expression,
$\mu_0\varepsilon\!\simeq\!(B_{c2}\!-\!B)/2.32\kappa^2$ for
$B\lesssim B_{c2}$,  with $\kappa\!=\!1.3$, and
$B_{c2}\!=\!0.29$T, we deduce from Eq.(\ref{two_mode}) the
$l_S(B)$-data  in Fig.\ref{slippage}. By taking the bulk
expression for $\varepsilon$, we disregard surface enhancement in
oblique field and therefore overestimate the $l_S^{-1}$ data in
this geometry. The high-field plateaus $l_S\!=\!l_0$ above
$B_{pk}$ in Fig.\ref{slippage} are suggestive of an individual
pinning, since in this case the slippage length is given by a
typical curvature radius of the corrugated surface, which should
not depend on the vortex density (magnetic field). This
interpretation is supported by the fact that the contact angle for
VL at the surface $\!a_0/l_0\!\simeq\!0.1$, estimated from
$l_0\!\simeq\!0.5\mu$m (normal field), agrees with the maximum
tilt $\zeta^*/a_0\!\sim 0.1$ deduced from  surface corrugation.
This suggests that the surface-pinning strength above $B_{pk}$ is
indeed limited by geometrical considerations. By contrast, the
strong suppression of $1/l_S$ below $B_{pk}$ (factor $\sim\!10$ in
oblique field) reflects the {\em collective} regime of surface
pinning, which  was known earlier in rotating $^3$He \cite{KKKS}.
The transition is sharp  unlike the continuous ones reported in
Refs.\cite{Paltiel00,Ling01}. Thus the experiment provides an
evidence that PE is accompanied by the crossover from the
collective to the individual surface pinning.

We start the theoretical analysis from calculation of the elastic
response of the semi-infinite vortex lattice to the surface force
presented by one Fourier component ${\mathbf f}({\mathbf r}) =
{\mathbf f}({\mathbf  k})e^{i {\mathbf k  r}}$. Any surface force
on a vortex should be balanced by the line-tension force:
${\mathbf f}({\mathbf r}) =\varepsilon\varphi_0
\partial {\mathbf u}/\partial z$. Thus the force produces the displacement
field in the bulk, which is assumed to be transverse ($\mathbf
U(\mathbf k) \perp \mathbf k$), since the VL is much more rigid
with respect to the longitudinal force than to the transverse one.
The equation of the elasticity theory is:
\begin{equation}
\left[C_{66} k^2 + C_{44} ( {\mathbf k}) k_z^2\right]{\mathbf
U}({\mathbf k})=0~,
   \label{EE}\end{equation}
where $C_{66}$ is the shear modulus and
\begin{equation}
C_{44}({\mathbf k}) = {B^2 \over \mu_0}\frac{1}{1 + \lambda^2(k^2
+k_z^2)} +{\varepsilon B}
   \end{equation}
is the tilt-modulus, which takes into account nonlocal effects due
to long-range vortex-vortex interaction. The solution of equation
(\ref{EE}) yields two values for $k_z^2=-p^2$, which correspond to
the {\em two modes} generated in the bulk. Usually $C_{66} \ll
\varepsilon B \ll B^2/\mu_0$ and approximately
\begin{equation}
p_1^2= \frac{(1+ \lambda^2k^2) C_{66} } {B^2/\mu_0 +\varepsilon B
\lambda^2 k^2}k^2~,~p_2^2={B\over \mu_0\varepsilon\lambda^2} + k^2
 \gg p_1^2~.
  \label{p1p2} \end{equation}
The solution is a superposition of two components, $U(x,z) = e^{ik
x}\left(u_1 e^{p_1 z} +u_2 e^{p_2 z}\right)$, with the amplitudes
$u_1$ and $u_2$ determined from the boundary conditions. The first
one is given by the surface force: ${ f}({\mathbf
k})=\varepsilon\varphi_0 (p_1 u_1 +p_2 u_2)$, and the second one
is that the tangential magnetic field determined by the London
equation, ${\mathbf h}({\mathbf k})=i k_z B {\mathbf U}({\mathbf
k})\left[1+\lambda^2( k^2 + k_z^2)\right]^{-1}$, vanishes at the
sample border. Eventually one can find the linear response
relation $f({\mathbf k})=C({ k}) U_0({\mathbf k})$, which connects
the force and the displacement $U(x,0)=U_0e^{ik x}$ at the surface
in the Fourier presentation, where $U_0=u_1 + u_2$ and
\begin{equation}
C(k) \approx k\varphi_0\sqrt{{C_{66}\over \mu_0} {(1+ \lambda
^2k^2 \mu_0\varepsilon/B)\over 1+\lambda ^2k^2}}~.
 \label{CdeK}  \end{equation}
Most of the deformation energy is stored in the first
long-wavelength mode.

The linear-response function allows to calculate the random
displacements produced by the surface pinning. Considering that
vortex cores end perpendicular to the local surface, the
corrugation produces a force $ f_m({\mathbf
r_i})=-\varepsilon\varphi_0{\partial \zeta({\mathbf
r_i})/{\partial {\mathbf x}_m}}$ applied to the end of the vortex
line, where ${\mathbf r}_i\!=\!{\mathbf R_i}+{\mathbf U}_0$ is the
2D position vector of the vortex end. For small displacements
${\mathbf U}_0$, the balance of force yields the boundary
condition
\begin{equation}
\varepsilon\varphi_0\left[{\partial {\mathbf U}_{0m}({\mathbf
R_i}) \over
\partial z}+\frac{\partial \zeta(\mathbf{R}_i)}{\partial
x_{\mathrm m}}\right]=0\quad.
                \label{force}\end{equation}
In the Fourier presentation the random force is ${\mathbf
f}({\mathbf  k}) =-i {\mathbf  k}\varepsilon\varphi_0  \int d
{\mathbf r}\, e^{-i {\mathbf k r }}\zeta(\mathbf r )$. Since $C(k)
\propto k$ at small $k$, the integral for the mean-square-root
displacement is divergent at small $k$. But we can obtain a
convergent integral for the mean-square-root shear deformation
\begin{eqnarray}
\left\langle{ ({\mathbf \nabla U}_0)^2}\right\rangle
=\left\langle{ (\partial U_{0x}/\partial y + \partial
U_{0y}/\partial x)^2}\right\rangle \nonumber \\= {\varepsilon
^2\varphi_0^2\over 4\pi ^2 }\int  {\mathbf k^2\,{\mathbf dk} \over
C(k)^2} \sum _{{\mathbf Q}}\left( Q^2 - { ({\mathbf k\cdot
Q})^2\over k^2}\right) S_\zeta({\mathbf k} +{\mathbf Q})
  \end{eqnarray}
Here the integration over $\mathbf k$ is fulfilled over the
Brillouin zone and $\mathbf Q$ is the reciprocal vortex-lattice
vector. We shall approximate the surface corrugation spectrum as
$S_\zeta({\mathbf k}) =2\pi \zeta^{*2} r_d^2e^{-kr_d} $, where
$\zeta^* =\sqrt{\langle \zeta({\mathbf R})^2\rangle_{\mathbf R} }$
is the amplitude of the surface roughness and $r_d$ is the
correlation length of the random pinning force, which cannot be
much less than the coherence length $\xi$. In the collective
regime $k\lesssim Q$, considering a broad corrugation spectrum
($r_d \ll a_0$),  summation over ${\mathbf Q} $ can be replaced by
integration, and we obtain
\begin{eqnarray}
\left\langle{ ({\mathbf \nabla U}_0)^2}\right\rangle =
{\varepsilon^2\varphi_0^2 a_0^2\over 8\pi  }\int
\limits_0^{2/a_0}} { k^3\,d{ k} \over C(k)^2} \int_0^\infty
S_\zeta({ Q)  Q^3\,dQ \nonumber \\ \approx{\varepsilon B\over
C_{66}}{3r_d^2 \over l_0^2}~,
  \label{def}\end{eqnarray}
where $l_0 = r_d^2/\zeta^*$ is on the order of the geometric
curvature radius. Here we have used the large k limit of
Eq.(\ref{CdeK}),  $C(k)\approx k\varphi_0(C_{66}\varepsilon/
B)^{1/2}$, which is a good approximation when $\lambda\gg a_0$.

In the experiment the AC fields produce additional quasistatic
displacements ${\mathbf u}({\mathbf{R}}_i)$ superimposed   on the
static displacements induced by pinning. By differentiating
Eq.(\ref{force}) for small ${\mathbf u}$, we obtain the boundary
condition:
\begin{equation}
{\partial {\mathbf u}_{\mathrm m}({\mathbf{R}}_i)  \over \partial
z} + \frac{\partial^2\zeta ({\mathbf{R}}_i)}{\partial x_{\mathrm
m}\partial x_{\mathrm n}}\;{\mathbf u}_{\mathrm n}({\mathbf R}_i)=
0\qquad. \label{rose}\end{equation} In the Fourier presentation
the uniform displacement (${\mathbf k} =0$) is coupled with the
nonuniform displacements (${\mathbf k} \neq 0$). Treating the
latter by the perturbation theory we arrive at the the boundary
condition Eq. (\ref{BC}) imposed on the averaged, i.e., uniform
displacement with slippage length given by
\begin{eqnarray}
\frac{1}{l_S}\simeq  {\varepsilon\varphi_0 \over  4\pi^2 }
\sum_{{\mathbf Q}}\int {\mathbf dk} \,|{\mathbf k}+ {\mathbf Q}|^2
\left[ Q^2 - { ({\mathbf k\cdot Q})^2\over k^2}\right] \nonumber
\\ \times {S_\zeta({\mathbf k}+ {\mathbf Q})\over C( k)}\quad.
\label{slippage_length}\end{eqnarray} Using the same approximation
as in calculating $\left\langle{ ({\mathbf \nabla
U}_0)^2}\right\rangle$ we receive for the inverse slippage length
\begin{eqnarray}
{1\over l_S} ={\varepsilon\varphi_0 a_0^2\over 8\pi } \int
_{0}^{2/a_0} {kdk \over C(k)}\int_0^\infty S_\zeta( Q)
Q^5\,dQ\approx\sqrt{\varepsilon B\over  C_{66}}{ 5! a_0\over 2
l_{0}^2}~.
  \label{lS}  \end{eqnarray}
Our derivation of the collective pinning strength $\propto 1/l_S$
was not based on the heuristic concept of Larkin-Ovchinnikov
domain \cite{Larkin} usually used for the analysis of collective
pinning.  The domain size $L_c$ is determined by the balance of
the pinning energy and the elastic energy. The average energy
density of pinning weakened by collective effects is $E_p \sim
\varepsilon\varphi_0 u^2/a_0^2l_0 \sqrt{N_c}$,  where $N_c
=L_c^2/a_0^2$ is the number of vortices in the Larkin-Ovchinnikov
domain. In order to estimate the elastic energy one should know
how far the shear deformation from surface disorder penetrates
into the bulk. This length is not evident {\em apriori}, but {\em
aposteriori}, from our analysis. Both experiment and theory
 show the importance of the short lengthscales,
$k^{-1} \sim a_0$ for the surface pinning; as a consequence,  the
deformation penetration length, $\sim 1/p_1 \sim
\sqrt{\varepsilon\varphi_0/C_{66}}$ which corresponds to large $k$
limit in Eq. (\ref{p1p2}), is finite. Then the elastic energy per
unit area is $E_{el} \sim (C_{66}/p_1) (u^2/L_c^2) \sim
\sqrt{C_{66} \varepsilon\varphi_0}u^2/L_c^2$. The condition $E_p
\sim E_{el}$ yields $L_c \sim l_0a_0\sqrt{C_{66}
/\varepsilon\varphi_0}$ and since $l_S \sim l_0\sqrt{N_c}=l_0
L_c/a_0$ one obtains for $l_S$ the value of the same order as in
Eq. (\ref{lS}).

Let us apply the received expressions for the magnetic fields far
from $B_{c2}$. In this case $\varepsilon \sim
(\varphi_0/\mu_0\lambda^2)\ln(B_{c2}/B)$, $C_{66} \sim \varphi_0
B/\mu_0\lambda^2$ and according to Eq. (\ref{lS}) $l_S \propto
\sqrt{B/\ln(B_{c2}/B)}$. Then the surface length $L_s \propto
[B/\ln(B_{c2}/B)]^ {3/2}$ grows with the magnetic field in
qualitative agreement with the experiment (Fig. 1). This is a
regime of collective pinning when $l_S > l_0$. At the same time
since $r_d \ll a_0$ the vortex lattice shear deformation remains
small according to Eq. (\ref{def}). Close to $B_{c2}$ $\varepsilon
\sim(B_{c2}-B)/\mu_0\kappa^2$, $C_{66} \sim
(B_{c2}-B)^2/\kappa^2$, and $r_d \sim a_0 \sim \xi$. Then from
Eqs. (\ref{def}) and (\ref{lS}) we receive that $\left\langle{
({\mathbf \nabla U}_0)^2}\right\rangle \approx
(\xi^2/l_0^2)B_{c2}/(B_{c2}-B)$  and  $l_S \approx (l_0^2/\xi)
\sqrt{(B_{c2}-B)/B_{c2}}$. Thus $l_S$ decreases when $B$
approaches to $B_{c2}$ and becomes smaller than $l_0$ for
$B_{c2}-B < B_{c2}\xi^2/l_0^2$. This means that surface pinning
ceases to be collective and the transition to the individual
pinning occurs. At the same time, at $B_{c2}-B \sim
B_{c2}\xi^2/l_0^2$, the deformation $\left\langle{ ({\mathbf
\nabla U}_0)^2}\right\rangle$ becomes of the order of unity, which
means destruction of the crystalline order at the surface. The
state without  the crystalline order near the surface can be
called {\em surface glass}. Thus the crossover from the collective
to the individual pinning is accompanied by the crystal-glass
transition at the surface.

Now let us check that this transition, or crossover, can explain
PE. Despite the fact that $l_S \propto \sqrt{B_{c2}-B}$ decreases
at $B$ approaching to $B_{c2}$, according to Eq. (\ref{two_mode})
$L_S$ continues to grow proportionally to $1/\sqrt{B_{c2}-B}$
whereas in the experiment (Fig. \ref{pic_effect}) $L_S(B)$
decreases close to the peak. However, the growth of the
deformation $\left\langle{ ({\mathbf \nabla
U}_0)^2}\right\rangle$, which accompanies the decrease of $l_S$
eventually invalidates the perturbative approach used above.
Qualitatively this can be corrected by introducing the
renormalized deformation-dependent shear modulus in the form
$\tilde C_{66}=C_{66}(1  - \alpha \left\langle{ ({\mathbf \nabla
U}_0)^2}\right\rangle) \approx C_{66}(1  -B/B_{pk})$. Here the
field $B_{pk}$ corresponds to the crystal-glass transition, where
$\tilde C_{66}=0$, and $\alpha$ is an unknown numerical factor,
which could be close to 0.1 as in the Lindemann criterion. Using
renormalized modulus $\tilde C_{66}$ in place of $C_{66}$ in Eq.
(\ref{lS}) we obtain that $l_S$ as well $L_S$  [see Eq.
(\ref{two_mode})] decrease inversely proportionally to
$\sqrt{B_{pk}-B}$ in qualitative agreement with  experiment (Figs.
\ref{pic_effect} and \ref{slippage}). On the right of the peak,
where pinning is individual and the shear rigidity is absent,
$l_S \sim l_0$ does not depend on the magnetic field $B$, while
$L_S \propto 1/(B_{c2}-B)$ grows with  $B$.

The close relation between PE and vanishing of the shear modulus
of VL was suggested in the early studies of PE
\cite{Pippard69,Larkin}. The new feature in our scenario is that
the shear vortex elasticity  vanishes only in some surface layer.
The width  $\sim 1/p_1 \propto 1/\sqrt{C_{66}}$ of this layer
diverges if $C_{66} \rightarrow 0$ and finally surface disorder
can contaminate the bulk of the sample. But this should happen for
fields $B> B_{pk}$, since the peak field $B_{pk}$ is determined by
$\tilde C_{66}$. Our scenario agrees with recent STM imaging of
the vortex array by Troyanovski {\em et al.} \cite{Troy}. This
experiment has revealed the disorder onset {\em on the surface} of
2H-NbSe$_2$ sample, coinciding with  the onset of the peak effect.
Troyanovski {\em et al.} related the disorder onset with the bulk
pinning. In order to discriminate two scenarios it would be useful
to supplement the STM probing of the vortex array at the surface
by probing vortex arrangements in the bulk.

In conclusion, this experiment and the accompanying theoretical
model provide a new plausible scenario for the peak effect. This
is a comprehensive demonstration of a new situation: the
order-disorder transition in a vortex lattice triggered by a weak
surface disorder. Beside its experimental relevance, this new
established mechanism offers an interesting paradigm for elastic
systems at the upper critical dimension for disorder.

We thank F.R. Ladan and E. Lacaze for technical assistance.  We
acknowledge fruitful discussions with B. Horowitz, T. Natterman
and T. Giamarchi. This work was funded by the French-Israel Keshet
program and by the grant of the Israel Academy of Sciences and
Humanities.

\newpage

\end{document}